\documentclass[3p,times,twocolumn]{elsarticle}

\usepackage{amssymb}

\usepackage[figuresright]{rotating}

\def\bea{\begin{eqnarray}}
\def\be{\begin{equation}}
\def\ee{\end{equation}}
\def\eea{\end{eqnarray}}

\def\b{\bigskip}
\def\m{\medskip}
\def\r{\rightarrow}

\begin{document}

\begin{frontmatter}




\title{What does strong subadditivity tell us about black holes?}


\author{Samir D. Mathur}

\address{Department of Physics,\\ The Ohio State University,\\ Columbus,
OH 43210, USA}

\begin{abstract}
It has been argued that small corrections to evolution arising from non-geometric effects can resolve the information paradox. We can get such effects, for example, from subleading saddle points in the Euclidean path integral. But an inequality derived in 2009 using strong sub-additivity showed that such corrections {\it cannot} solve the problem. As a result we sharpen the original Hawking puzzle: we must either have (A) new (nonlocal) physics  or (B) construct hair at the horizon. We get correspondingly  different approaches to resolving the AMPS puzzle. Traditional complementarity assumes (A); here we require that the AMPS experiment measures the correct vacuum  entanglement of Hawking modes, and invoke nonlocal $A=R_B$ type effects to obtain unitarity of radiation. Fuzzball complementarity is in category (B); here the AMPS measurement is outside the validity of the approximation required to obtain the complementary description, and a effective regular horizon arises only for freely infalling observers with energies $E\gg T$.

\end{abstract}


\end{frontmatter}


\section{Hawking's puzzle}
\label{fail}

\begin{figure}[htbp]
\begin{center}
\includegraphics[scale=.38]{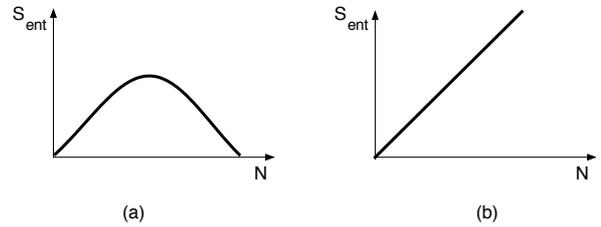}
\caption{{(a) For a normal body, the entanglement of the body with its radiation $S_{ent}$ at first rises, and then falls back to zero. (b) For a black hole which has the Unruh vacuum at its horizon, $S_{ent}$ keeps rising monotonically.}}
\label{fonedel}
\end{center}
\end{figure}

The no-hair  `theorem'  tells us that black holes are featureless; described by the Schwarzschild metric
\be
ds^2=-(1-{2M\over r})dt^2+{dr^2\over 1-{2M\over r}}+r^2d\Omega^2
\label{one}
\ee
Hawking \cite{hawking}  showed that such a hole radiates by producing particle pairs. One member of the pair, in a mode we label  $b$, escapes to infinity.  The other member $c$ falls into the hole. The crucial point is that $b$, $c$ are in an entangled state that we may write schematically as
\be
|\Psi\rangle_{pair}={1\over \sqrt{2}}\big ( \, |0\rangle_b|0\rangle_c+|1\rangle_b|1\rangle_c\big )
\label{three}
\ee
The entanglement entropy $S_{ent}$ between the inside and outside of the hole thus keeps growing, with the entanglement after $N$ steps being
\be
S_{ent}=N\ln 2
\label{two}
\ee
This is in sharp contrast to the behavior of a normal body, where $S_{ent}$ starts to reduce after the halfway point and reaches zero at the end of the evaporation process \cite{page}. For the black hole, the result (\ref{two}) creates difficulties  near the endpoint of evaporation, a problem referred to as the black hole information paradox.  Fig.\ref{fonedel} depicts these different behaviors of $S_{ent}$.

\section{An attempt at a resolution: the idea of `approximate emergent spacetime'}\label{sectwo}

String theory aims to yield  a unitary theory of gravity, but not all string theorists were overly worried by Hawking's argument. The idea of AdS/CFT duality \cite{adscft} seemed to provide a reason: if gravity was dual to a unitary gauge theory, then how can there be a problem with black hole radiation?

It is true that amplitudes for low energy processes in gravity  were reproduced by the dual gauge theory. But when the energy reached the black hole threshold, people just wrote down the AdS-Schwarzschild metric to describe the state. This was natural, since the no-hair `theorem' seems to allow no other gravity solution. But Hawking's argument can be applied to this metric just as it was applied to the Schwarzschild metric (\ref{one}). So the relativists were puzzled: won't we again run into  Hawking's problem of monotonically rising entanglement?

In answer to this criticism, many string theorists believed the following: ``The black hole is really described by a field theory with some number $n$ of unitarily evolving degrees of freedom, for example the gluons of the CFT description. In the gravity description, spacetime slices etc. are only an effective, approximate emergent construct. Thus Hawking's computation, which used `evolution using good slices across the horizon', would be only approximately correct. In particular, Hawking's computation seems to enlarge the effective Hilbert space by the creation of new pairs. But the true description involves a  fixed number of bits $n$, so the created quanta must be some complicated combinations of  the already existing bits. When we take such effects into account, we should find that the radiation from the hole is unitary, with an entanglement graph like fig.\ref{fonedel}(a).''

For concreteness, we summarize this belief in the following model of black hole evaporation:

\b

(i) For $r>10  M$ we assume that spacetime is flat with normal, local physics. Quanta that escape to this region are not modified further as they travel out to infinity. (This is also what happens for the unitary evolution of a piece of burning paper; photons that escape the paper do not suffer any modification large enough to affect the unitarity question.)

\m

\begin{figure}[htbp]
\begin{center}
\includegraphics[scale=.38]{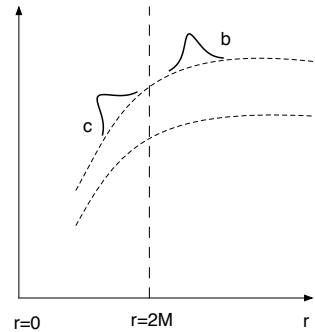}
\caption{{Creation of a pair in the approximate spacetime that emerges from an $n$-bit system that is supposed to model the black hole. The local vacuum on the lower slice evolves to contain an entangled pair on the later slice.}}
\label{ftwodel}
\end{center}
\end{figure}

(ii) In the region $r<10 M$ we write the traditional metric of the hole, but assume that this spacetime is only an effective, approximate concept. We can look at one smooth spacelike slice through the horizon, and study its evolution to the next slice where one $b,c$ pair is created (fig.\ref{ftwodel}). Very nonperturbative processes, perhaps involving 
geometries other than the black hole, can contribute small corrections to this evolution.

\m

(iii) The details of pair creation for {\it one} pair process will be close to that given by semiclassical physics, since otherwise there would be no meaning at all to the black hole metric that we have written down. Thus we create a $b,c$ pair at this step of evolution, and the properties of these quanta should approximate the properties expected from the semiclassical evolution.

\m

(iv) There is however  no requirement that we describe accurately the creation of a {\it large} number of Hawking pairs using such a semiclassical slicing. The error in the semiclassical description of each pair is small, but these errors can cumulate when we try to write the state for a large number of pairs.

\m

(v) The $b$ quantum escapes to the region $r>10M$, representing the Hawking radiation at this step. This radiation process is claimed to have the behavior appropriate to a normal body, with $S_{ent}$ given by fig.\ref{fonedel}(a) and not fig.\ref{fonedel}(b). 

\b

If such a model were true, then Hawking's puzzle would be a non-puzzle from the start. After all, there can always be small subleading corrections to a leading order computation. If such corrections could alter the graph of fig.\ref{fonedel}(b) to fig.\ref{fonedel}(a), then there would be no reason to be overly concerned by the fact  that Hawking's leading order computation gave the monotonically rising entanglement (\ref{two}). In fact Hawking, in 2004, agreed that there was a probably a resolution to his original puzzle along similar lines \cite{hawkingreverse}. He noted an argument (related to one by Maldacena \cite{eternal}) where the small corrections could be attributed to the effect of subleading saddle points in the path integral.  

\section{Failure of this approach: the strong subadditivity argument}

But in 2009, an inequality was derived, using strong sub-additivity, which showed that any such model {\it cannot} work \cite{cern,avery,acp}. In other words, the inequality showed that no unitary model can approximate the black hole spacetime well enough to describe approximately even {\it one} step of Hawking pair creation. In outline, the proof proceeds as follows:

\b

(a) Let the quanta emitted in emission steps $1, 2, \dots N$ be denoted $\{ b_1, b_2, \dots b_N\} \equiv \{ b\}$. The entanglement of the radiation with the hole at step $N$ is then
\be
S_N=S(\{b\})
\ee
where $S(A)$ for any set $A$ denotes the entanglement of $A$ with the remainder of the system. 

\m

(b) The bits in the hole evolve to create an `effective bit' $b_{N+1}$ and an `effective bit' $c_{N+1}$. (The bit $b_{N+1}$ has not yet left the region $r<10M$.) The entanglement of the earlier emitted quanta $\{ b\}$ does not change in this evolution. (If two parts of a system are entangled, and we make a unitary rotation on one part, the entanglement between the parts does not change.) 

\m

(c) The effective bits $b_{N+1}, c_{N+1}$ must approximate the properties of the Hawking pair (\ref{three}). In (\ref{three}) we have $S(b_{N+1}, c_{N+1})=0$, since the pair is not entangled with anything else. We also have $S(c_{N+1})=\ln 2$. Thus for our model we must have 
\be
S(b_{N+1}+c_{n+1})<\epsilon_1
\label{four}
\ee
\be
S(c_{N+1})>\ln 2 -\epsilon_2
\label{five}
\ee
for some $\epsilon_1\ll1$, $\epsilon_2\ll 1$. 

\m

(d) The bit $b_{N+1}$ now moves out to the region $r>10M$. The value of $S_{ent}$ at timestep $N+1$ is
\be
S_{N+1}=S(\{ b\}+b_{N+1})
\label{six}
\ee
since now $b_{N+1}$ has joined the earlier quanta $\{ b\}$ in the outer region $r>10M$. 

\m

(e) We now recall the strong subadditivity relation
\be
S(A+B)+S(B+C)\ge S(A)+S(C)
\ee
We wish to set $A=\{ b\}$, $B=b_{n+1}$, $C=c_{N+1}$. We note that these sets are made of independent bits:
(i) The quanta $\{ b \} $ have already left the hole and are far away (ii) The quantum $b_{n+1}$ is composed of some bits, but as it moves out to the region $r>10M$, it is independent of the bits remaining in the hole and also the bits $\{ b \} $ (iii) The quantum $c_{N+1}$ is made of bits which are left back in the hole. Applying the strong subadditivity relation, we get
\bea
S(\{ b_i\}+b_{N+1})+S(b_{N+1}+c_{N+1})&\ge&\cr
 S(\{ b_i \})&+&S(c_{N+1}) 
\label{seven}
\eea
Using (\ref{four}),(\ref{five}),(\ref{six}) we get
\be
S_{N+1}> S_N+\ln 2 -(\epsilon_1+\epsilon_2)
\label{eight}
\ee
Thus for $\epsilon_1, \epsilon_2\ll 1$, the entanglement keeps growing in the manner of fig.\ref{fonedel}(b) and cannot behave like
that of a normal body (fig.\ref{fonedel}(a)). 

\b

Thus we conclude that  having an `approximate emergent space-time' instead of the smooth space-time used in Hawking's original calculation \cite{hawking} does not resolve the information paradox.

\section{Consequences of the inequality (\ref{eight})}\label{secfour}

The inequality (\ref{eight}) has very important consequences for the information paradox:

\b

(a)  {\it Hawking's argument of 2004:}
 In 2004 Hawking \cite{hawkingreverse} had suggested that  subleading saddle points in the path integral can contribute corrections that would remove the entanglement between quanta inside and outside the hole. But we now see that small corrections {\it cannot} remove the entanglement, even if we allow these corrections to arise from effects that do not appear geometric in the original black hole space-time. 

\b

(b) {\it Arguments using AdS/CFT:} Many people had come to believe that the idea of AdS/CFT duality implied that information had to emerge in the Hawking radiation. But (\ref{eight}) shows  that this reasoning is not correct. Below the black hole threshold, CFT correlators are observed to agree with gravity correlators. But the information paradox does not arise below the black hole threshold. Above the threshold, people just wrote down the AdS-Schwarzschild metric for the hole, which has the same problem of rising entanglement as the Schwarzschild hole. So how have we resolved the problem?

People thought that the problem would be resolved because AdS/CFT gave only an approximation to the full AdS-Schwarzschild geometry. Leading order physics of pair creation would have to be reproduced in this approximation, because there is no meaning to writing the AdS-Schwarzschild metric if we cannot use it even for the simple low energy process of creating {\it one} pair. But there could be small corrections from  {\it non-geometric} effects at each step of pair creation, since space-time was only an effective emergent notion. It could then be hoped that the non-geometric corrections would make $S_{ent}$ behave like fig.\ref{one}(a) rather than fig.\ref{one}(b).

But the inequality (\ref{eight}) shows that this hope is false; $S_{ent}$ would continue to grow as in fig.\ref{one}(a), if the AdS-Schwarzschild metric is able to describe approximately even one step of pair creation, regardless of what non-perturbative effects are used to generate the corrections. 

\b

In fact (\ref{eight}) sharpens the original Hawking paradox, which arises from a combination of two results:

\m

(i) The no-hair `theorem': any quanta near the horizon get sucked into the hole, so it is hard to get any state other than the Unruh vacuum around the horizon.

\m

(ii) Hawking's 1975 computation \cite{hawking} shows that the Unruh vacuum leads to pair production with the growing entanglement (\ref{two}).

\b

The result (\ref{eight}) then says that we have two sharply different choices, which we label (A) and (B):

\b

(A) We accept the conclusions of the no-hair `theorem', and assume that the vacuum at the horizon is the Unruh vacuum. If we then want to avoid the consequences of the growing entanglement (\ref{two}), then  we must have some {\it new} physics. For example:

\m

(1) We postulate the different observers see different things. For example the approach we will call `traditional complementarity' postulated that observers who stay outside the hole and observers who fall inside do not see states in  the same Hilbert space, since they cannot compare observations before encountering the black hole singularity \cite{complementarity}. 

\m

(2) We postulate that there is a `final state boundary condition' at the black hole singularity. The singularity appears in the `future' of the evolution on the nice slices of the black hole, so we lose the usual notions of causality and locality in time.   Such a model was proposed in \cite{hm}, and has recently been analyzed in \cite{preskill}. 

\m

(3) We postulate that high energy physics is local, but low energy modes have some non-local behavior which modifies the process of particle creation. Such effects have been postulated for example by Giddings \cite{giddings}.

\m

(4) We modify assumption (i) of the model in section \ref{sectwo}; i.e., we allow quanta that have escaped the hole to be modified by nonlocal effects even after they have gone very far. Since Hawking evaporation of a solar mass hole takes $\sim 10^{77}$ years, this means that we are invoking nonlocal effects over distances $\sim 10^{77}$ light years. The model of \cite{raju} attempts to evade the inequality (\ref{eight}) by effects of this form (though the fact that they need such nonlocal effects is not made very clear). 

\m

(5) We postulate that when two distant systems are entangled, space-time is modified so that a wormhole connects them. Such a postulate was made recently by Maldacena and Susskind \cite{cool}. 

\b

(B)  We find a way around the no-hair `theorem'. This requires us to construct structure at the horizon that `stays there' instead of falling in and restoring the state back to the local vacuum. If the state at the horizon is not the Unruh vacuum, then Hawking's computation of pair creation does not apply, and we are not forced to have the rising entanglement (\ref{two}). Such hair has been found in string theory by the fuzzball construction \cite{fuzzballs}.\footnote{The construction was motivated by  solutions found earlier in \cite{earlier}.}

\b

  \begin{figure}[htbp]
\begin{center}
\includegraphics[scale=.38]{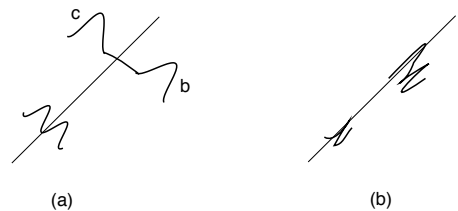}
\caption{{(a) Evolution of modes at the horizon, in a situation where the state at the horizon is given in some approximation  by the vacuum (b) If we find hair however, then the evolution at the `would-be' horizon can be quite different, and we need not get the formation of entangled pairs.}}
\label{ffivedel}
\end{center}
\end{figure}

The difference between (A) and (B) is illustrated schematically in fig.\ref{ffivedel}. Fig.\ref{ffivedel}(a) depicts the physics of (A), where we have not found any way to construct hair. Thus we have the standard black hole geometry across the horizon, and we have the creation of an entangled pair $(b,c)$ in some approximation. Fig.\ref{ffivedel}(b) shows a situation where we {\it have} managed to construct hair; i.e., a state which does not settle does to the Unruh  vacuum at the horizon. Now the modes involved in the Hawking process evolve in a completely different way, and we do {\it not} need to have the creation of entangled pairs $(b,c)$. 

\section{Avoiding a fallacy: `the circular argument'}

Before discussing in  more detail the difference between options (A) and (B), we wish to note a common confusion that can blur the sharp distinction between these two options. Some people say: ``I believe in AdS/CFT, but I don't trust the AdS-Schwarzschild metric as the correct one inside the hole. So I don't have the nice slices used in the derivation of entangled $(b,c)$ pairs, and so am not concerned with the problem created by monotonically rising entanglement.''

But such a statement makes no sense. The information paradox arises from the very fact that people were not able to find a wavefunctional for gravity which resembled the hole outside the horizon but did not automatically continue as the vacuum wavefunctional  across the horizon (the no-hair `theorem'). So if someone wanted to avoid having the vacuum at the horizon then he has to {\it show} a construction of hair which achieves this, and this is difficult because any quanta placed near the horizon tend to fall in and restore the local state back to the vacuum. Thus if someone says that he does not trust the traditional hole in the interior, but does not  provide a mechanism for `hair', then he is avoiding the question that is basic to the information paradox: {\it how} do we get something other than the vacuum at the horizon? One should not {\it assume} that there will be hair -- the whole problem is to find a {\it construction} of hair. 

This same fallacy often appears in a slightly different form. Some people say: ``If I define the gravity theory as dual to a unitary CFT, then won't I {\it have} to get the information out in the Hawking radiation? And if that is so, then don't I know that there will be hair, even though I do not know the construction of this hair?''

But this argument is incorrect as well, for a reason which can be seen by considering the different timescales that arise in black hole physics.\footnote{Full details can be found in \cite{not}; here we summarize the argument.} For a Schwarzschild hole in 3+1 dimensions,  the crossing timescale is $\sim M$. The Hawking evaporation timescale is $\sim M^3$. If the hole releases information more slowly than this (say, over times $\sim M^4$) then we can say we have a `remnant'. The dual to the CFT must be unitary, but this fact does not tell us which of these three possibilities we get:

\m

(i) The $c=1$ matrix model provides a toy model of AdS/CFT duality. But here we find that the energy in a collapsing shell bounces back in a time which is the analogue of the crossing time $\sim M$; thus no black hole forms \cite{pol}. This time is also the most `natural' timescale in the CFT: below the black hole threshold, quanta thrown in towards the center of AdS return back in `crossing time'. What happens is that the CFT agrees with expectations of 1+1 gravity at low energies, but stringy corrections modify this expectation  when the energy is increased to the black hole threshold.

\m

(ii) A similar situation holds for other CFTs. Low energy amplitudes agree between the CFT and gravity. But stringy corrections can create large changes by the time we get to the black hole threshold. The energy of an infalling shell may be trapped only for times analogous to $\sim M$, as for the 1+1 d case; in that case there would be no black hole. If a larger trapping time emerges, then this time could be, say $\sim M^4$, in which case we would have a remnant.

\b

(iii) We could hope that the trapping time be $\sim M^3$, so that the gravity description contains the analog of a black hole evaporating at the Hawking rate. But in this situation we are faced with Hawking's puzzle: if we cannot construct a gravity solution that has hair, then we cannot have the information emerge in the Hawking radiation.

\b

Thus we see that if we define the gravity theory as being the dual of the CFT, then we are no closer to understanding the behavior of black holes: we are still left with the whole range of possibilities, from no black hole formation all the way to remnants.  

The discovery of AdS/CFT duality was exciting since it added a lot to our understanding of gravity. For the first time we had a picture that could encompass all intricacies of quantum gravity, packaged as a CFT that could be completely solved in principle.  But in this excitement many people arrived at an erroneous conclusion that the black hole information paradox had been resolved as well. The thinking seemed to run: (i) There is  a unitary dual (ii) The most natural situation would be that information emerge in Hawking radiation (iii) Since spacetime is only an approximate construct in this scheme, corrections to Hawking' computation would remove the problem of ever increasing entanglement. 

But with the inequality (\ref{eight}) we see that step (iii) is incorrect.  We are thus forced to confront Hawking's problem in exactly the same terms as before the advent of AdS/CFT: we must choose one of the two options (A) and (B) in section \ref{secfour}, and find the physics needed to back up our choice.

\section{The fuzzball construction}

The fuzzball construction  attains option (B) of section \ref{secfour}, by showing how a structure can be constructed at the horizon which will not `fall in' and return the horizon to the vacuum state. String theory has extra dimensions; let us assume that these are compactified to circles. Normally we expect that these compact directions are trivially tensored with the noncompact ones, and only serve to provide the low energy particle content of the theory. But near the black hole horizon these compact directions can `pinch-off'  providing an {\it end} to space-time. The fluxes, branes etc of the theory all appear in the detailed solution to support this pinch-off structure. But the essential point is that this structure cannot `fall in' through the horizon, because the structure itself  provides a way for space-time to end outside the horizon.   

Radiation emerges from the surface of the fuzzball just as it would emerge from the surface of a piece of coal. Thus the radiation is {\it not} produced by the process of pair creation at a horizon. {\it Thus we do not recover the production of even a simple pair $(b,c)$ in any approximation.} We may get some effective dynamics of the fuzzball when we consider `high energy impacts' (fuzzball complementarity), but it is crucial to note that this effective dynamics will never be accurate enough to describe the evolution of the modes involved in the Hawking process. 

In short, the fuzzball construction says that the solutions of string theory at energies above the black hole threshold live in a different topological class: in this class the outer region ends without boundary outside the horizon. This situation gives a structure that does not `fall in' and restore the Unruh vacuum at the horizon. 

We can now conjecture what happens to a shell that is collapsing towards its horizon radius. Consider this motion using Schwarzschild time $t$. The shell appears to slow down as it approaches its horizon radius $r=2M$.  At the same time another process starts to happen: the wavefunction of the shell starts to spread over the entire phase space of fuzzball solutions, which have the same quantum numbers as the shell. This spread is rapid because the number of fuzzball states making this phase space is very large -- the number of states is given by $Exp[S_{bek}]$. Thus at late times $t$ we do not get a shell that reaches $r=2M$; instead we get a complicated superposition of fuzzball eigenstates \cite{tunnel}. These fuzzballs radiate from their surface like a normal body, so we do not have the creation of entangled pairs, and thus no Hawking paradox. 

  \begin{figure}[htbp]
\begin{center}
\includegraphics[scale=.38]{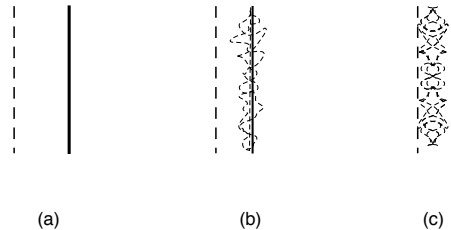}
\caption{{(a) A collapsing shell approaches its horizon (the dotted line) (b) a part of the wavefunction of the shell spreads over the space of fuzzball states (c) after a long time, the wavefunction has moved out of the shell state and spread completely over the space of fuzzball states.}}
\label{fsixdel}
\end{center}
\end{figure}

The fuzzball proposal been confusing to many people for different reasons, so we turn next towards clarifying these issues.

\section{Common questions about fuzzballs, and their answers}

Let us explain the nature of the fuzzball construction by going over some common questions:

\b

(a) {\it Are fuzzballs classical or quantum?} They are certainly quantum. Even for the spacetime in this room, the metric $ds^2=\eta_{ab}dx^adx^b$ is only an approximation; what we have really is a wavefunctional $\Psi[{}^{(3)}g, \phi]$ over 3-geometries and any matter fields $\phi$. Similarly, the fuzzball states should really be thought of as wavefunctionals $\Psi[{}^{(9)}g, \Phi]$ where $\Phi$ represents schematically all the nongravitational fields of string theory.  The crucial point is that the manifolds ${}^{(9)}g$ appearing in the wavefunctional are in a different topological class from the manifolds that were traditionally assumed to appear -- the manifolds ${}^{(9)}g$ in $\Psi$ end compactly before forming a horizon, while in normal vacuum state at the horizon they would continue smoothly through into the black hole interior.  This compactness is achieved because the compact directions are not trivially tensored with the noncompact ones. 

\b

(b) {\it The generic black hole microstate is expected to have curvatures of order planck scale. How can we hope to understand anything about  about this structure since any description will be hopelessly messy?} The crucial point is that we can arrange the states in a sequence of `complexity'. The simplest states in this sequence can be explicitly seen to be described by gravitational solutions that end compactly outside the horizon, and then we can extrapolate to all states. 

The situation is analogous to how people looked for `hair' in the early days of black hole physics. For a scalar field  one would write the ansatz $\Phi=f(r)Y_{l,m}(\theta, \phi)e^{-i\omega t}$ and look for solutions $f(r)$ for each angular harmonic $l=0, 1, 2, \dots$. It turned out there were no hair in this ansatz, but had we actually found valid solutions $f(r)$ then there would have been no Hawking puzzle. The solutions for low $l$ would be reliably known. We would then extrapolate to the cutoff $l_{max}\sim R/l_p$ where $R$ is the horizon radius; this would give  $\sim (R/l_p)^2$ `hair modes', the correct order to match the Bekenstein entropy.  The modes for $l\sim R/l_p$ would be the generic ones, but they would also be `messy' since all quantum gravity effects would have been involved at wavelengths $\lambda\sim l_p$. Nevertheless, if we had found valid solutions $f(r)$ for $l=0,1, 2, \dots$, we would assume that the generic hole had a horizon which carried its data, instead of having a horizon which was the local {\it vacuum}.

The fuzzball construction achieves what the perturbative ansatz $\Phi=f(r)Y_{l,m}(\theta, \phi)e^{-i\omega t}$ failed to do: it arranges all states in a sequence of complexity (using the dual CFT as a guide) and then finds explicit constructions for the `hair' starting from the simplest states. The generic state is indeed expected to be a quantum mess, but what is crucial is that the states constructed involve only solutions that end compactly outside the horizon. From this  we conjecture that all the states of the hole should be in this new topological class.

\b

(c) {\it If the generic fuzzball is messy and quantum, and has structure at the planck scale, then  is it distinguishable from the vacuum, which is also messy and quantum and has structure at the planck scale?} The answer is  {\it yes}; the fuzzball state is indeed distinguishable from the vacuum. In fact the generic fuzzball state is nowhere near the vacuum which leads to Hawking pair production. Understanding the answer to this question is crucial, since it has led to most of the confusions about fuzzballs, including (a) and (b) above. 

\begin{figure}[htbp]
\begin{center}
\includegraphics[scale=.38]{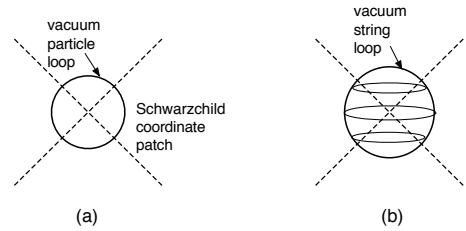}
\caption{{(a) A virtual loop for a scalar particle; in the right wedge covered by Schwarzschild coordinates, it appears that particles bubble in and out of the horizon (b) (b) The same situation for a string loop; now it appears that a string gas lives near the horizon.}}
\label{fthreedel}
\end{center}
\end{figure}

The source of this confusion can be traced back to the formulation of complementarity in the early nineties. At that time, {\it no} construction of hair had been found. Thus on a `good slice' through the horizon, one had to assume the local Unruh vacuum state. To avoid the information problem, {\it new} physics was postulated in the form of an {\it observer dependence} of states. In fig.\ref{fthreedel}(a) we depict spacetime seen in Kruskal coordinates. This spacetime is locally a vacuum, with the rightwards sloping null line giving the position of the horizon.  The circular line depicts a vacuum bubble: a pair of scalar quanta emerge from the vacuum, live for a while and then annihilate.  Such fluctuations are of course part of the natural physics of the local vacuum. 

Now consider the Schwarzschild coordinates for this situation, which cover only the right wedge of the diagram. In these coordinates it appears that a particle emerged from the horizon, and then fell back in. Quite generally, it appears that the Unruh vacuum is `bubbling with quanta' when viewed in the Schwarzschld frame. This bubbling, in turn can be traced to the fact that the Schwarzschild coordinates cover only a part of the full manifold, and at the place where they break down we have $g_{tt}\r 0$.  Similarly, we can consider the vacuum bubble of a string (fig.\ref{fthreedel}(b)); now in Schwarzschild coordinates it appears that strings emerge and disappear at the horizon. 

One can now postulate: 

\m

{\it Traditional complementarity, assumption (i):}  These excitations visible in the Schwarzschild frame will absorb and re-radiate quanta that fall onto the Schwarzschild hole, acting as a `stretched horizon' which scrambles and returns information back to infinity {\it for the purposes of an outside observer}.

\m

But a relativist would have an immediate objection: these excitations are just vacuum fluctuations seen in bad coordinates, and when the effects of all these fluctuations is summed over, an infalling quantum should continue to move unimpeded through the vacuum state at the horizon. We thus postulate 

\m

{\it Traditional complementarity, assumption (ii):}  For the purposes of an infalling observer, the physics is indeed that of smooth infall through a vacuum region.

\m

Now we are faced with a conflict: the `no-cloning theorem' tells us that we cannot have the information reflected to infinity by the stretched horizon and also have the information fall in to the interior of the horizon. Thus we postulate:

\m

{\it Traditional complementarity, assumption (iii):} The postulates (i) and (ii) are not in conflict because the observations of an observer who stays outside should not be thought of as living in the same Hilbert space as the observations of an observer who falls in. It is true that both observers can be captured on the same Cauchy slice through spacetime, but there is not enough time for these two observers to compare notes before the singularity is reached, and in this situation a single Hilbert space should not describe all the data on the Cauchy slice.

\m

To summarize, in traditional complementarity we do not have any construction of hair.  We evade the information problem by postulating that the Hilbert space of states is not defined using global Cauchy slices; instead we have observer dependent Hilbert spaces and observables.

Now let us contrast this with the situation with `fuzzball complementarity', where we do start with an explicit construction of hair:

\m

{\it Fuzzball complementarity, postulate (i):} The states of the hole are given by wavefunctionals $\Psi[{}^{(9)}g, \Phi]$ where the ${}^{(9)}g$ are manifolds in a different topological class from the Unruh vacuum: they end compactly outside the horizon due to a pinch-off of one or more compact cycles. Infalling quanta are absorbed by these `real' degrees of freedom at the horizon, and re-radiated as Hawking radiation.

\m

{\it Fuzzball complementarity, postulate (ii):} The physics of the modes $b_i,c_i$ is not reproduced in any approximation. Thus we do not recover the Unruh vacuum at the horizon.

\m

\begin{figure}[htbp]
\begin{center}
\includegraphics[scale=.38]{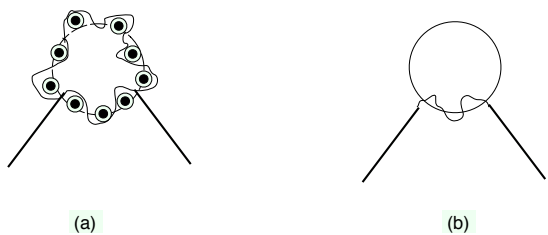}
\caption{{(a) Probing the fuzzball with operators at energy $E\gg kT$ causes collective excitations of the fuzzball surface. (b) The corresponding correlators are reproduced in a thermodynamic approximation by the traditinal black hole geometry, where we have no fuzzball structure but we use the geometry on both sides of the horizon.}}
\label{feetwo}
\end{center}
\end{figure}

 {\it Fuzzball complementarity, postulate (iii):} If the fuzzball surface is impacted hard by freely infalling\footnote{Here and in what follows, `freely infalling' implies that the object is dropped in from a distance $r-R\gtrsim R$. This is in contrast to lowering the object gently to a position close to the horizon, and then dropping it in. In particular, measurements made by observers who `hover' near the horizon are excluded by the the `freely infalling' requirement.}  quanta with $E\gg T$, then to a good {\it approximation} we get collective oscillations of this fuzzball surface. The Green's functions defining this collective dynamics can be obtained using the traditional black hole metric which has no fuzzball structure but does have the region interior to the horizon (fig.\ref{feetwo}).
 
 \m
 
 To summarize, there is {\it no} observer dependence of physics. The structure of hair is covariant, and not coordinate dependent. Thus we call the fuzzball structure as giving `real degrees of freedom' as opposed to the `virtual degrees of freedom' that one obtains by viewing the Unruh vacuum in Schwarzschild coordinates.

\b

 Now we can return to the confusion that we are trying to clarify. There are {\it two} completely different ways in which one may encounter monopoles near the horizon:
 
 \m

\begin{figure}[htbp]
\begin{center}
\includegraphics[scale=.38]{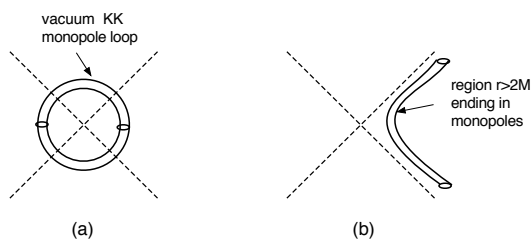}
\caption{{(a)  A vacuum loop formed by a KK monopole (b) A completely different situation, where the spacetime in the right wedge {\it ends} in KK monopoles.}}
\label{ffourdel}
\end{center}
\end{figure}

 (i)  If we have the Unruh vacuum at the horizon, then we would have virtual loops of monopoles, just as we had virtual loops of scalar quanta and string in fig.\ref{fthreedel}(a),(b).  We depict such a vacuum loop for the monopole in  fig.\ref{ffourdel}(a).
 
 \m
 
 (ii) By contrast, in a fuzzball solution, the spacetime {\it ends} just outside the horizon 
 by the pinch-off of compact directions. This pinch-off creates a monopole structure, depicted in fig.\ref{ffourdel}(b), but this is {\it not} the structure of the vacuum loop of fig.\ref{ffourdel}(a). 
 
  \m
  
 It is important to not confuse these two appearances of the monopoles with each other. In (i) we just have the natural fluctuations of the vacuum, and an infalling quantum should not get reflected back from these fluctuations (unless we postulate new physics). By contrast in (ii),  spacetime  ends at these monopoles. Thus an infalling particle cannot really `pass through' to the black hole interior, and an `effective interior' is realized in a very different way through the notion of fuzzball complementarity.
 
To summarize, the fuzzball solutions will indeed become more messy and quantum as we move to generic fuzzballs, but this should not make us confuse vacuum fluctuations (fig.\ref{ffourdel}(a)) with real degrees of freedom (fig.\ref{ffourdel}(b)). In particular, vacuum fluctuations like those of fig.\ref{fthreedel}(a) are present even for a theory of canonical gravity
plus scalar fields, but there is {\it no} fuzzball structure for this case. Nor do we get fuzzball structure by just considering strings moving in the Schwarzschild geometry (\ref{one}). Fuzzballs are a particular nonperturbative construction that require many aspects of string theory.\footnote{Ref. (\cite{gibbonswarner}) explains how this construction evades the assumptions that were implicit in many derivations of the no-hair theorems.} With fuzzballs, the production of entangled Hawking pairs $(b,c)$ is not reproduced to {\it any} approximation. Only {\it high impact} ($E\gg T$) processes have an approximate description in terms of free infall through the horizon.

\section{The AMPS argument}

The difference between traditional complementarily and fuzzball complementarity can be seen quite explicitly in the different ways they addresses the recent AMPS argument \cite{amps} which was formulated tho exclude traditional      complementarity. 

The analysis of \cite{cern} showed that a regular horizon implies rising entanglement. Equivalently, we can say that if entanglement is to decrease, then the state at the horizon cannot be the vacuum. AMPS adapted this analysis  to suggest a crisp and elegant argument against complementarity \cite{amps}, which we summarize as follows. Let Picture 1 describe the physics of an `outside observer', and Picture 2 the physics of an observer who falls in smoothly through the horizon.

\b

(a) Consider Picture 1. The radiation $\{ b_i\}$ from earlier steps of emission is near infinity. The quantum that has just been emitted, $b_{N+1}$ is outside, but close to the stretched horizon. 

\b

(b) Now consider Picture 2. We denote quanta in this picture with a prime $'$. It is assumed that everything outside the stretched horizon is identical between the two pictures:
\be
\{ b'_i\}~=~\{ b_i\}
\label{bppq}
\ee
\be
b'_{N+1}~=~b_{N+1} \,.
\label{bbp}
\ee


(c) In Picture 2 we assume that we have the vacuum at the horizon. Thus the mode across the horizon $c'_{N+1}$ is entangled with $b'_{N+1}$ in the manner assumed in Hawking's computation. Thus (\ref{eight}) gives
 \be
  S'_{N+1} ~\gtrsim~ S'_N+\ln 2 \,.
  \ee
 (AMPS ignore the small corrections, setting $\epsilon_1, \epsilon_2=0$.) 
 
  \b
  
  (d) In Picture 2 we were not looking for unitarity of evaporation, since the infalling observer did not have time to measure the entanglement of emitted quanta. But by (\ref{bppq}), (\ref{bbp}) we have $S'_N=S_N$, $S'_{N+1}=S_{N+1}$. Thus we find that, in Picture 1,
  \be
  S_{N+1}\gtrsim S_N+\ln 2
  \ee
  This contradicts the fact that in Picture 1 we do want the entanglement to {\it decrease}, after the halfway point, by approximately $\ln 2$ per emitted bit.  Thus AMPS argue, we cannot have complementarity.
  
  \b

The resolution of the AMPS puzzle in fuzzball complementarity was given in \cite{mt2}. There have also been several papers addressing the issue in  traditional complementarity,  for example \cite{cool} which postulated that entangled states should be thought of as joined by wormholes. Let us now note the differences between these different approaches to the AMPS puzzle:

\b

(1) {\it Measurement of the quantum $b_{N+1}$:} In traditional complementarity, it is postulated that the quantum $b_{n+1}$ can be measured and will give the same result in pictures 1 and 2. 

In fuzzball complementarity, picture 1 is exact while picture 2 is only an approximation good for reproducing certain measurements (those appropriate to normal physics obtained in a freely falling frame). It is shown (sec.5.1 of \cite{mt2}) that  for an observer that falls freely from afar towards the horizon, the measurement of $b_{N+1}$ is a transplanckian process, not covered in the approximation of fuzzball complementarity. More precisely,  if a freely infalling observer tries to switch on and off a detector quickly enough to measure $b_{N+1}$, then he will get burnt by the observation himself as the time needed for this switching is shorter than planck time.  Thus we should not try to cover measurements of $b_{N+1}$ in picture 2. Thus eq.(\ref{bbp}) does not hold. 

\b

(2) {\it The role of $c_{N+1}$:} In traditional complementarity we want to get the local vacuum in picture 2, so we ask that an infalling observer be able to measure the mode $c_{N+1}$, and check its entanglement with the measurable mode $b_{N+1}$.

In fuzzball complementarity, there is no interior to the fuzzball surface in the exact description of picture 1. In the approximate description of picture 2, no analogue of $c_{N+1}$ emerges, since the physics of $(b,c)$ modes is not captured in the approximation inherent in picture 2.

\b

(3) {\it Hovering vs. free infall:} The AMPS argument did not make a distinction between measurements that could be made by a freely falling observer and an observer who is fine-tuned to hover near the horizon. Traditional complementarity makes no such distinction either, since all observations outside the stretched horizon are assumed to me possible and to agree between pictures 1 and 2. 

In fuzzball complementarity only the observations natural to a freely falling observer are expected to be well approximated by pictures 2, which mimics free infall through a smooth horizon. If an observer is fine-tuned to hover within some planck lengths of the horizon, then his observations need not be well approximated by picture 2. 

\b

(4) {\it The condition $E\gg T$:} AMPS make no use of any energy condition like $E\gg T$ on the observations that are supposed to have a complementary description. Traditional complementarity has not had any such condition either. 

In fuzzball complementarity, on the other hand, the black hole interior obtained in picture 2 is only obtained in an approximation that gets better as $E/T\r\infty$. The breakdown of the approximation for the $E\sim T$ hawking modes is what allows the information to escape in the Hawking radiation.

It is important to avoid the following confusion. Consider an observer falling into the hole. In his frame, quanta of energy $E\sim T$ have a wavelength $\lambda\sim R$, where $R$ is the radius of the hole. Thus these quanta do not {\it fit} into the black hole, and so there is no sense in asking for the dynamics of $E\lesssim T$ quanta in the complementary description of picture 2. We could make this observation even for traditional complementarity, so one might think that the condition $E\gg T$  is somehow implicit in traditional complementarity as well.  

But such is not the case;    condition $E\gg T$ appearing in fuzzball complementarity has quite a different implication. In fuzzball complementarity the Hawking modes $b,c$ {\it emerging} from the horizon fail to be reproduced by the effective geometry of picture 2. (The $b$ modes have $E\sim T$ when they reach infinity.) Getting an approximation where these $(b,c)$ modes are not captured is very nontrivial: it requires the space-time to first be altered completely at the horizon (which the fuzzball construction accomplishes), and then for the effective dynamics of the fuzzball surface to give rise to an emergent interior.  We {\it cannot} get to this situation without first being able to construct `hair'.  It is also true that the accuracy of picture 2 for {\it infall}  improves as $E/T\r \infty$, but for the information problem it is the issue of {\it outgoing} modes $b,c$ that is crucial.

\b

(5) {\it The mechanism for obtaining complementarity:}  In traditional complementarity we have the vacuum at the horizon in picture 2. Most of the approaches to resolving the AMPS puzzle have been based on postulating a nonlocal effect that  identifies Hawking radiation modes at infinity with modes inside the horizon. Such approaches have been loosely termed $A=R_B$, where $A$ are the modes in the hole and $R_B$ are the Hawking radiation modes at infinity. For example in \cite{cool} it is postulated that a wormhole structure connects this radiation not the interior of the horizon. 

In fuzzball complementarity we have no such nonlocal identification of modes; in fact no new physics is postulated and the fuzzball surface radiates just like a piece of burning paper. The mechanisms \cite{mt2} that are relevant to resolving the AMPS puzzle involve (i) the fact that the stretched horizon moves outwards {\it before} it is impacted by the in falling quantum (ii) the impact by a quantum with $E\gg T$ creates states that are mostly {\it new}, so they are not entangled with the radiation at infinity\footnote{The observation that new states are involved is also involved in the discussion of \cite{ver}.} (iii) it is the dynamics of these new states that is captured by picture 2 that mimics free infall through a smooth horizon. None of these mechanisms appear in the approaches involving traditional complementarity.

\b

\begin{figure}[htbp]
\begin{center}
\includegraphics[scale=.38]{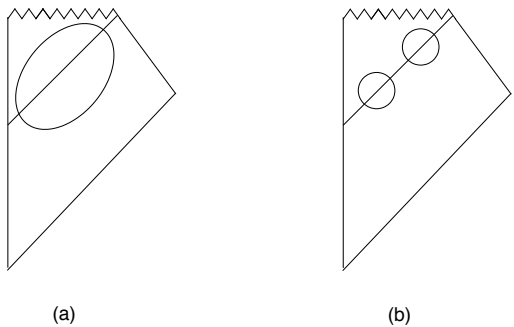}
\caption{{(a) In traditional complementarity nothing constrains us from having the domain marked by the ellipse as the region where the complementary description applies. (b) In fuzzball complementarity, the complementary description is good only with the small patches indicated by circles; the approximation involved in complementarity starts to fail if we try to join such small patches into line large patch.}}
\label{fdel10}
\end{center}
\end{figure}

(6) {\it Domain of complementary description:} In traditional complementarity there is no explicit restriction on how much of the hole is covered by the complementary description. In fact the entire region outside the stretched horizon is described by ordinary physics, and so can be covered in one patch. There is no claim that we cannot cover the entire interior of the hole as well in one patch, since we expect the state to be the vacuum for every observer falling across the horizon  (see fig.\ref{fdel10}(a)). 

In fuzzball complementarity, we can describe only small regions at a time with the complementary description. The condition $E\gg T$ restricts us to consider space-time patches that are of size $L<< T^{-1}\sim M$. Thus we can cover one of the patches in fig.\ref{fdel10}(b), but if we try to join our description of several such patches together, then there will be a mismatch in the overall region. This mismatch will be in modes of the type involved in the Hawking process; in fact this mismatch is what allows Hawking radiation to carry information in the complementary description.

\b

(7) {\it Physical effects:} The above differences in the resolutions of the AMPS puzzle also lead to a difference in the eventual physics effect that we get for infall into black holes. Consider a black hole which is maximally  entangled with its radiation  $R$. Now suppose we consider an server who jumps into the black hole. For traditional complementarity, it was a argued in \cite{cool} that the physics felt by this observer can be altered by making a unitary transformation $R\r UR$ on the far away radiation. In particular, whether this observer sees a smooth horizon or a firewall would depend on the operation $U$.

By contrast, in fuzzball complementarity the black hole and its  radiation are just ordinary quantum systems that happen to be entangled. We  cannot change any physical effects concerning the hole by operating on the far away radiation $R$. 

\b

To summarize, traditional complementarity has virtual degrees of freedom at the horizon; since there is no construction of hair, the state at the horizon is the local vacuum. Fuzzball complementarity starts with a construction of hair, so there are real degrees of freedom at the horizon, which cause space-time to `end' there. The interior of the hole is then obtained only as an effective approximation, where the modes $(b,c)$ involved in the Hawking process do not survive the approximation. The consequences of this difference are very large, as can be seen from the very different ways that they address the AMPS argument. 

\section{Conclusion}

The inequality (\ref{eight}) derived using strong subadditivity has three important consequences for back holes:

\b

(a) Some people had the belief that the idea of AdS/CFT duality removed the need to worry about the information paradox. The idea was that since the black hole spacetime is only an approximate description of a unitary underlying CFT, nonperturbative effects that are non-geometric in the original spacetime would be able to reduce the growing entanglement (\ref{two}). The inequality (\ref{eight}) shows that this belief is false; {\it no} small corrections can solve the problem, and we are back to square one on the information problem. Thus we must either have new physics (option (A) of sec.\ref{secfour}) or find a way to construct hair (option (B) of sec.\ref{secfour}). 

\b

(b) The fuzzball construction realized option (B) by giving a construction of hair. But some people believed that this construction described only a special subclass of states; generic states of the hole would behave effectively like the vacuum near the horizon. The inequality (\ref{eight}) shows however that if we are to realize option (B), then {\it all} states of the hole need to be fuzzballs. If some states did reduce to the vacuum for the purposes of low energy modes, then we could not remove the entanglement problem (\ref{two}) by any set of small corrections. We would then need to invoke `new physics', some examples of which were listed in sec.\ref{secfour}.

\b

(c) The fuzzball construction led to the notion of fuzzball complementarity. Here the dynamics of the modes $(b,c)$ involved in the Hawking process is {\it not} reproduced in the complementary description, but the physics of observers who fall freely onto the fuzzball surface {\it can} be given an approximate complementary description that mimics infall through a smooth horizon. By contrast traditional complementarity requires the state at the horizon to be the vacuum, so that the $b,c$ modes involved in the Hawking process should exhibit the entanglement required from a local vacuum. This leads to a very different set of approaches to resolving the AMPS puzzle. In fuzzball complementarity we invoke no new physics, while in traditional complementarity most of the proposed solutions invoke nonlocal $A=R_B$ type approaches that postulate a nonlocal identification between radiation modes far away and the interior of the hole.

\b

\b

{\it Acknowledgements:} This work was supported in part by DOE grant DE-FG02-91ER-40690. I thank Steve Avery, Borun Chowdhury and  especially David Turton for many helpful comments. I thank the organizers of Light Cone 2012 for inviting me to talk. 




\nocite{*}
\bibliographystyle{elsarticle-num}
\bibliography{martin}



\end{document}